\documentstyle[twoside,psfig]{article}

\catcode`\@=11
\long\def\@makefntext#1{
\protect\noindent \hbox to 3.2pt {\hskip-.9pt  
$^{{\eightrm\@thefnmark}}$\hfil}#1\hfill}		

\def\@makefnmark{\hbox to 0pt{$^{\@thefnmark}$\hss}}	
	
\def\ps@myheadings{\let\@mkboth\@gobbletwo
\def\@oddhead{\hbox{}
\rightmark\hfil\eightrm\thepage}   
\def\@oddfoot{}\def\@evenhead{\eightrm\thepage\hfil
\leftmark\hbox{}}\def\@evenfoot{}
\def\sectionmark##1{}\def\subsectionmark##1{}}

\oddsidemargin=\evensidemargin
\newcounter{sectionc}\newcounter{subsectionc}\newcounter{subsubsectionc}
\renewcommand{\section}[1] {\vspace{12pt}\addtocounter{sectionc}{1} 
\setcounter{subsectionc}{0}\setcounter{subsubsectionc}{0}\noindent 
	{\tenbf\thesectionc. #1}\par\vspace{5pt}}
\renewcommand{\subsection}[1] {\vspace{12pt}\addtocounter{subsectionc}{1} 
	\setcounter{subsubsectionc}{0}\noindent 
	{\bf\thesectionc.\thesubsectionc. {\kern1pt \bfit #1}}\par\vspace{5pt}}
\renewcommand{\subsubsection}[1] {\vspace{12pt}\addtocounter{subsubsectionc}{1}
	\noindent{\tenrm\thesectionc.\thesubsectionc.\thesubsubsectionc.
	{\kern1pt \tenit #1}}\par\vspace{5pt}}
\newcommand{\nonumsection}[1] {\vspace{12pt}\noindent{\tenbf #1}
	\par\vspace{5pt}}

\newcounter{appendixc}
\newcounter{subappendixc}[appendixc]
\newcounter{subsubappendixc}[subappendixc]
\renewcommand{\thesubappendixc}{\Alph{appendixc}.\arabic{subappendixc}}
\renewcommand{\thesubsubappendixc}
	{\Alph{appendixc}.\arabic{subappendixc}.\arabic{subsubappendixc}}

\renewcommand{\appendix}[1] {\vspace{12pt}
        \refstepcounter{appendixc}
        \setcounter{figure}{0}
        \setcounter{table}{0}
        \setcounter{lemma}{0}
        \setcounter{theorem}{0}
        \setcounter{corollary}{0}
        \setcounter{definition}{0}
        \setcounter{equation}{0}
        \renewcommand{\thefigure}{\Alph{appendixc}.\arabic{figure}}
        \renewcommand{\thetable}{\Alph{appendixc}.\arabic{table}}
        \renewcommand{\theappendixc}{\Alph{appendixc}}
        \renewcommand{\thelemma}{\Alph{appendixc}.\arabic{lemma}}
        \renewcommand{\thetheorem}{\Alph{appendixc}.\arabic{theorem}}
        \renewcommand{\thedefinition}{\Alph{appendixc}.\arabic{definition}}
        \renewcommand{\thecorollary}{\Alph{appendixc}.\arabic{corollary}}
        \renewcommand{\theequation}{\Alph{appendixc}.\arabic{equation}}
        \noindent{\tenbf Appendix \theappendixc #1}\par\vspace{5pt}}
\newcommand{\subappendix}[1] {\vspace{12pt}
        \refstepcounter{subappendixc}
        \noindent{\bf Appendix \thesubappendixc. {\kern1pt \bfit #1}}
	\par\vspace{5pt}}
\newcommand{\subsubappendix}[1] {\vspace{12pt}
        \refstepcounter{subsubappendixc}
        \noindent{\rm Appendix \thesubsubappendixc. {\kern1pt \tenit #1}}
	\par\vspace{5pt}}

\newcommand{\textlineskip}{\baselineskip=13pt}
\newcommand{\smalllineskip}{\baselineskip=10pt}

\def\eightcirc{
\begin{picture}(0,0)
\put(4.4,1.8){\circle{6.5}}
\end{picture}}
\def\eightcopyright{\eightcirc\kern2.7pt\hbox{\eightrm c}} 

\newcommand{\copyrightheading}[1]
	{\vspace*{-2.5cm}\smalllineskip{\flushleft
	{\footnotesize International Journal of Modern Physics A, #1}\\
	{\footnotesize $\eightcopyright$\, World Scientific Publishing
	 Company}\\
	 }}

\def\abstracts#1#2#3{{
	\centering{\begin{minipage}{4.5in}\baselineskip=10pt\footnotesize
	\parindent=0pt #1\par 
	\parindent=15pt #2\par
	\parindent=15pt #3
	\end{minipage}}\par}} 

\newcommand{\bibit}{\nineit}

\renewenvironment{thebibliography}[1]
	{\frenchspacing
	 \ninerm\baselineskip=11pt
	 \begin{list}{\arabic{enumi}.}
	{\usecounter{enumi}\setlength{\parsep}{0pt}
	 \setlength{\leftmargin 12.7pt}{\rightmargin 0pt} 
	 \setlength{\itemsep}{0pt} \settowidth
	{\labelwidth}{#1.}\sloppy}}{\end{list}}

\newcounter{itemlistc}
\newcounter{romanlistc}
\newcounter{alphlistc}
\newcounter{arabiclistc}

\newcommand{\fcaption}[1]{
        \refstepcounter{figure}
        \setbox\@tempboxa = \hbox{\footnotesize Fig.~\thefigure. #1}
        \ifdim \wd\@tempboxa > 5in
           {\begin{center}
        \parbox{5in}{\footnotesize\smalllineskip Fig.~\thefigure. #1}
            \end{center}}
        \else
             {\begin{center}
             {\footnotesize Fig.~\thefigure. #1}
              \end{center}}
        \fi}

\newcommand{\tcaption}[1]{
        \refstepcounter{table}
        \setbox\@tempboxa = \hbox{\footnotesize Table~\thetable. #1}
        \ifdim \wd\@tempboxa > 5in
           {\begin{center}
        \parbox{5in}{\footnotesize\smalllineskip Table~\thetable. #1}
            \end{center}}
        \else
             {\begin{center}
             {\footnotesize Table~\thetable. #1}
              \end{center}}
        \fi}

\def\@citex[#1]#2{\if@filesw\immediate\write\@auxout
	{\string\citation{#2}}\fi
\def\@citea{}\@cite{\@for\@citeb:=#2\do
	{\@citea\def\@citea{,}\@ifundefined
	{b@\@citeb}{{\bf ?}\@warning
	{Citation `\@citeb' on page \thepage \space undefined}}
	{\csname b@\@citeb\endcsname}}}{#1}}

\newif\if@cghi
\def\cite{\@cghitrue\@ifnextchar [{\@tempswatrue
	\@citex}{\@tempswafalse\@citex[]}}
\def\citelow{\@cghifalse\@ifnextchar [{\@tempswatrue
	\@citex}{\@tempswafalse\@citex[]}}
\def\@cite#1#2{{$\null^{#1}$\if@tempswa\typeout
	{IJCGA warning: optional citation argument 
	ignored: `#2'} \fi}}

\def\pmb#1{\setbox0=\hbox{#1}
	\kern-.025em\copy0\kern-\wd0
	\kern.05em\copy0\kern-\wd0
	\kern-.025em\raise.0433em\box0}

\def\fnt#1#2{\footnotetext{\kern-.3em
	{$^{\mbox{\scriptsize #1}}$}{#2}}}

\def\fpage#1{\begingroup
\voffset=.3in
\thispagestyle{empty}\begin{table}[b]\centerline{\footnotesize #1}
	\end{table}\endgroup}

\font\tenrm=cmr10
\font\tenit=cmti10 
\font\tenbf=cmbx10
\font\bfit=cmbxti10 at 10pt
\font\ninerm=cmr9
\font\nineit=cmti9

\font\eightrm=cmr8

\def\qed{\hbox{${\vcenter{\vbox{			
   \hrule height 0.4pt\hbox{\vrule width 0.4pt height 6pt
   \kern5pt\vrule width 0.4pt}\hrule height 0.4pt}}}$}}

\begin{document}


\normalsize\textlineskip
\thispagestyle{empty}
\setcounter{page}{1}

\copyrightheading{}			

\vspace*{0.88truein}

\fpage{1}
\centerline{\bf TESTING PERTURBATIVE RESULTS WITH}
\vspace*{0.035truein}
\centerline{\bf NON-PERTURBATIVE METHODS FOR}
\vspace*{0.035truein}
\centerline{\bf THE HIERARCHICAL MODEL}
\vspace*{0.37truein}
\centerline{\footnotesize Y. MEURICE and M. B. OKTAY
\footnote{Presenting Author}}
\vspace*{0.015truein}
\centerline{\footnotesize\it The University of Iowa, Department of Physics and
Astronomy}
\baselineskip=10pt
\centerline{\footnotesize\it Iowa City, Iowa 52242, USA }

\vspace*{0.21truein}
\abstracts{We present non-perturbative methods to calculate accurately
the renormalized quantities for Dyson's Hierarchical Model. We apply
this method and calculate the critical exponent $\gamma$ with 12 and 4
significant digits in the high and low temperature phases, respectively.
We report accurate values for universal ratios of amplitudes and 
preliminary results concerning the comparison with perturbative results.}{}{}


\vspace*{1pt}\textlineskip	
\section{Introduction}	
\vspace*{-0.5pt}
The scalar field theory has many important applications such as the
generation of mass in the standard model of elementary particles via 
spontaneous symmetry breaking and the theory of critical phenomena.
However, there is no approximate treatment of the theory which can compete
in accuracy with the perturbative treatment of quantum electrodynamics.


Hierarchical approximations, where the Renormalization Group (RG) 
transformations can be performed easily, provide a test ground for obtaining 
accurate calculations as well as for testing the validity of several
perturbative techniques such as the renormalized perturbative expansion 
in field theory. Here, we present a approximation where simple algebraic 
methods can be used to calculate the renormalized quantities with great 
accuracy.

We consider Dyson's Hierarchical Model\cite{dyson} that couples the main
spin in boxes of size $2^l$ with a strength $(c/4)^l$ where $c$ is a free 
parameter set to $c=2^{1-2/D}$ in order to approximate a nearest neighbor
model in $D$ dimensions. We have to specify a local measure $W_0(\phi)$,
for instance of the Landau-Ginsburg type $W_0(\phi)\propto e^{-[(1/2)m^2\phi^2
+\lambda_B\phi^4]}$ where the bare parameters will appear, or of the Ising 
type. Under a block spin transformation, the local measure changes according
to 
\begin{equation}
W_{n+1}(\phi)\propto e^{{\beta\over2}({c\over4})^{n+1}}\times
\int d\phi^{`} W_n({\phi^{`}-\phi\over2})W_n({\phi^{`}+\phi\over2})
\end{equation}
The recursion formula can be re-expressed in Fourier representation as
\begin{equation}
R_{n+1}\propto {\rm exp}({-1\over2}\beta{\partial^2\over\partial k^2})
(R_n({\sqrt{c}k\over2}))^2
\end{equation}
where $R_n(k)$ is the Fourier transform of $W_n(\phi)$ with a rescaling of
$\phi$ by a factor $2c^{-1/2}$.

\section{Calculational Method and Numerical Results}

It was found that the finite dimensional truncations of degree $l_{max}$ :
$R_n(k)=1+a_{n,1}k^2+a_{n,2}k^4+...+a_{n,l_{max}}k^{2l_{max}}$ provide
very accurate results in the symmetric phase\cite{godina1,godina2}. The 
round-off errors on the two point function grow like $(\beta-\beta_c)^{-1}$
and the finite truncations and volume effects can be controlled with 
an exponential precision. The non-Gaussian distribution of the numerical
errors made in RG calculations related to small changes in the scaling factor
was also studied in great detail\cite{meurice1}. The critical exponent 
associated with the two point function was found 
$\gamma=1.299140730159\mp10^{-12}$. It was also shown that hyperscaling holds
in $D=3$ and that the coupling constant vanishes like $1/\Lambda$ in $D=4$.

In the broken symmetry phase, it is possible to take advantage 
of about 10 iterations for which the low temperature scaling is
 observed\cite{godina3} to extrapolate to infinite volume limit. 
The critical exponents corresponding to zero momentum connected $q-$point
functions $G_q^c(0)$, for $q=1,2,3$ are $\gamma_1=-0.324775\mp2\times10^{-5}$
, $\gamma_2=1.29918\mp1-^{-4}$ and $\gamma_3=2.928\mp10^{-2}$, respectively,
in good agreement with the hyperscaling relation\cite{godina3}
\begin{equation}
\gamma_q=\gamma[(q/2){\rm ln}(4/c)-{\rm ln}2]/{\rm ln}(2/c).
\end{equation}

Recently, we have observed that the dimensionless renormalized 
coupling constants in the high temperature phase
\begin{equation}
\lambda_q^\star={G_{2q}^c\;m_R^{q(2+D)-D}\over \beta^{(q-1)D/2}}
\end{equation}
corresponding to the $q-$point function tend to universal values independent
of the choice of the initial measure. Here $G_{2q}^c$ is the connected Green
function and $m_R$ is the renormalized mass. We list below the values with
significant digits common to an Ising and a Landau-Ginsburg measures.

\begin{table}[htbp]
\tcaption{Renormalized coupling constants 
$\lambda_q^\star$ for $q-$ point functions.}
\centerline{\footnotesize\smalllineskip
\begin{tabular}{cccc} \\
\hline 
$q$ & $\lambda_q^\star$ & $q$ & $\lambda_q^\star$\\
\hline 
$4$ & $1.5058$ & $8$ & $579.97$ \\
$6$ & $18.107$ & $10$ & $35653$ \\
\hline
\end{tabular}}
\end{table}

\section{Perturbative Analysis}

We will use the accurate results mentioned above to test perturbation
theory. We expand the initial local measure into series as

\begin{equation}
e^{-(1/2)m^2\phi^2-\lambda_B\phi^4} \simeq 
e^{-(1/2)m^2\phi^2}(1-\lambda_B\phi^4+{\lambda_B^2\over2}\phi^8+...) .
\end{equation}

\begin{figure}
\vspace*{13pt}
\centerline{ 
\psfig{width=6.8cm,figure=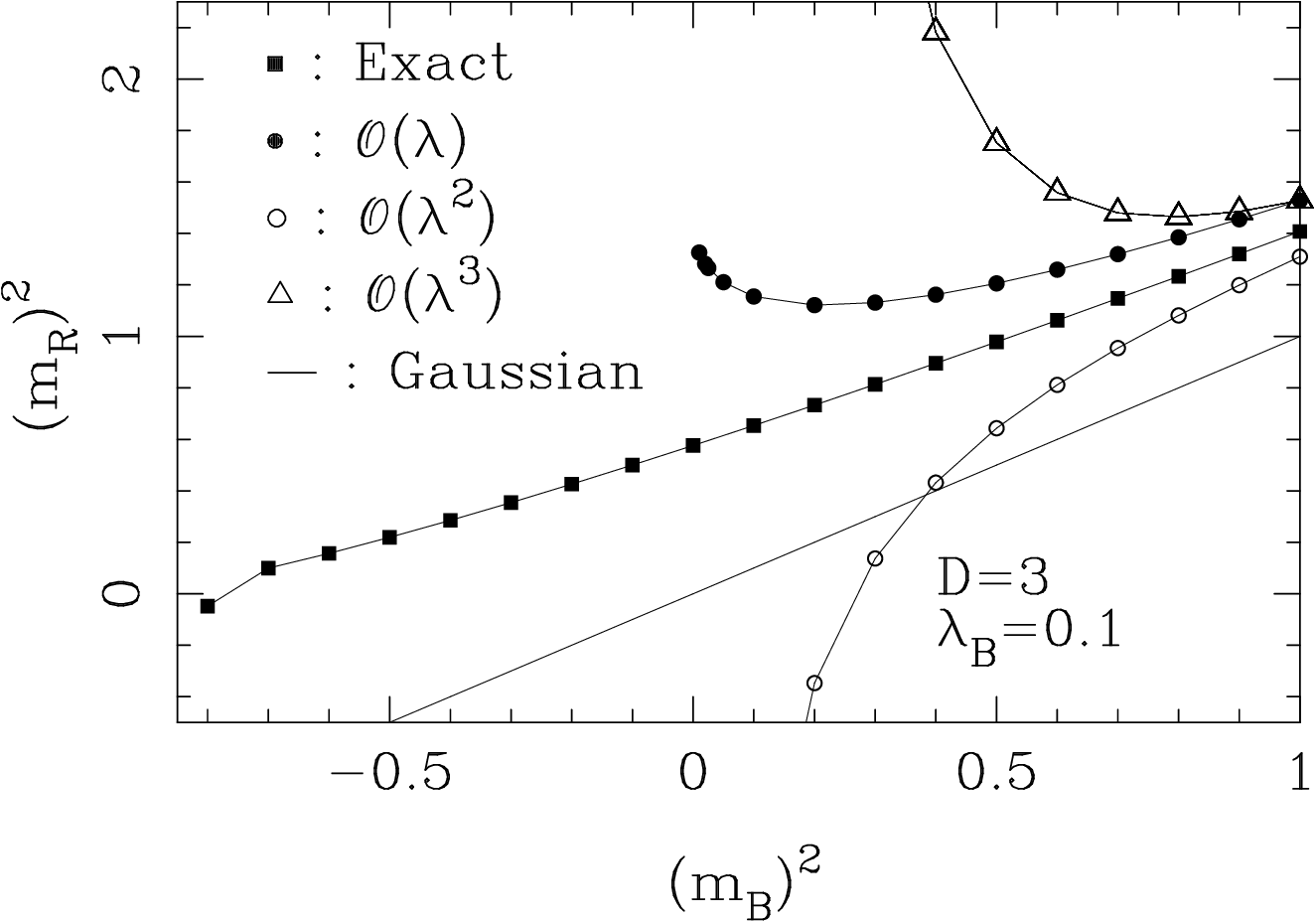,angle=270}
\psfig{width=6.8cm,figure=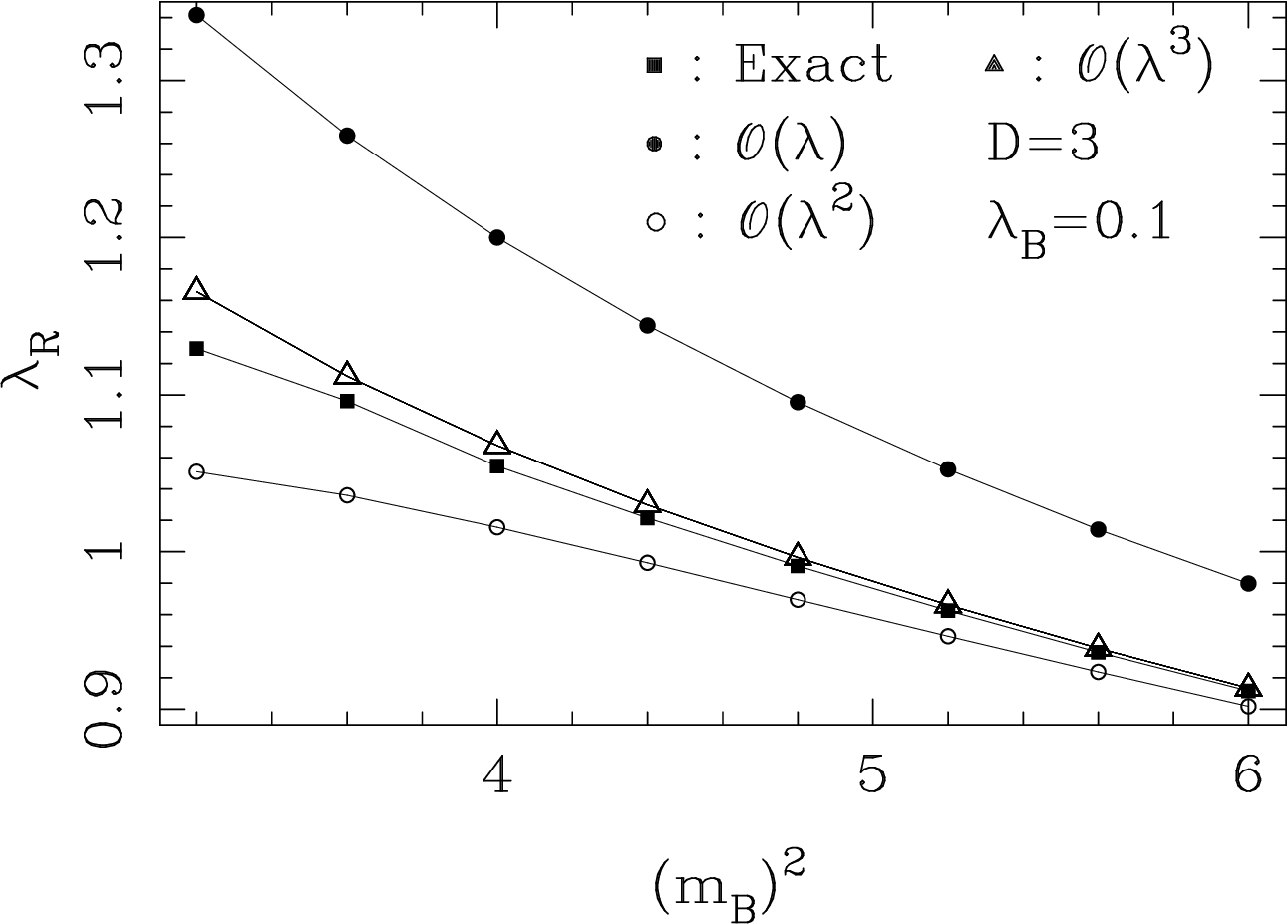,angle=270}
}
\fcaption{ (a) Dimensionless Renormalized mass versus dimensionless bare mass;
(b) Dimensionless Renormalized Coupling Constant versus dimensionless bare
 mass.}
\end{figure}

Substituting this back into the recursion relation, one obtains the recursion
relation in orders of $\lambda_B$. After $n$ iterations, the local measure
becomes
\begin{equation}
W_n(\phi)={\rm exp}[-{(m_B^2+k^2(n))\over 2^{n+1}\phi^2}]\times(1-\lambda_B
\times({3\over2^n}\phi^2\sum_{m=0}^{n-1}{2^{-m}\over (m_B^2+k^2(m))}+
{\phi^4\over 2^{3n}})+\vartheta(\lambda_B^2)
\end{equation}

\noindent
where $m_B^2$ stands for the bare mass and $k^2(l)$ can be interpreted as the 
momentum square over $l-$iterations. We have calculated the renormalized
mass and the dimensionless coupling constant up to order of $\lambda_B^3$.
The details of the results will be provided soon\cite{meurice2}. This 
is illustrated in the figures above. The iterative method can be used to 
high order in $\lambda_B$ if one does not insist on obtaining closed form
as in Eq. (6). This will be used to test standard perturbation
 method\cite{parisi} to evaluate the critical exponents.

\nonumsection{References}

\end{document}